\documentclass[twocolumn,showpacs,preprintnumbers,amsmath,amssymb]{revtex4}

\usepackage{dcolumn}
\usepackage{bm}
\usepackage{amsmath}
\usepackage{epsfig}             
\usepackage{epstopdf}         
\usepackage{hyperref}          
\usepackage{color}
\usepackage{colortbl}
\usepackage{graphicx}
\usepackage{amssymb,amsmath}
\usepackage{subfigure}
\usepackage{enumerate}
\usepackage{gensymb}

\begin{document}

\title{Signatures of double-electron re-combination in high-order harmonic generation driven by spatially inhomogeneous fields}

\author{Alexis Chac\'on$^1$}
\author{Marcelo F. Ciappina$^2$}
\author{Maciej Lewenstein$^{1,3}$}

\affiliation{$^1$ICFO-Institut de Ciencies Fotoniques, The Barcelona Institute of Science and Technology, 08860 Castelldefels (Barcelona), Spain}
\affiliation{$^2$Max-Planck-Institut f\"ur Quantenoptik, Hans-Kopfermann-Strasse 1, D-85748 Garching, Germany}
\affiliation{$^3$ICREA-Instituci\'o Catalana de Recerca i Estudis Avan\c{c}ats, 08010 Barcelona, Spain}

\begin{abstract}
We present theoretical studies of high-order harmonic generation (HHG) driven by
plasmonic fields in {\em two-electron} atomic systems. Comparing the two-active electron and single-active electron approximation models of the negative hydrogen ion atom, we provide strong evidence that a double non-sequential two-electron recombination
appears to be the main responsible for the HHG cutoff extension.
Our analysis is carried out by means of a reduced one-dimensional numerical integration of the
 two-electron time-dependent Schr\"odinger equation (TDSE),
and on investigations of the classical electron trajectories
resulting from the Newton's equation of motion. Additional comparisons
between the negative hydrogen ion and the helium atom suggest that
the double recombination process depends distinctly on the atomic target. Our
research paves the way to the understanding of strong field
processes in multi-electronic systems driven by spatially
inhomogeneous fields.
\end{abstract}

\pacs{42.65.Ky, 32.80.Rm, 33.20.Xx, 32.80 Qk, 42.50 Ct}

\maketitle

The incessant development of ultrafast, femtosecond ($10^{-15}$~fs)
laser technology in the infrared (IR) regime opened new avenues to
study a wide range of strong-field laser matter processes at their
natural time scale~\cite{MNisoli1997,AHuillier1993,Krausz:2009zz}.
These invaluable experimental and technological tools allowed
physicists to address instrumental aspects of one of the most
fundamental processes: the tunneling ionization of atoms and
molecules~\cite{Krausz:2009zz}. In particular, the application of
this laser technology provided a key factor for the  understanding
of the main mechanisms underlying the emission of coherent
radiation from atoms or
molecules~\cite{Ferray1988,AHuillier1993,Corkum1993,LewensteinPRA1994}.

As a matter of fact, one could say the high-order harmonic generation
(HHG) process fits within the tunneling ionization regime, when the
Keldysh parameter, defined by $\gamma=\sqrt{\frac{I_p}{2U_p}}$, is
$\gamma\leq1$. $I_p$ and $U_p=\frac{\mathcal{E}_0^2}{4\omega_0^2}$
denote here the ionization potential
 of the atomic target, and the electron ponderomotive potential energy in atomic
 units, respectively.
$\mathcal{E}_0$ is the peak amplitude of the laser electric field
and $\omega_0$ the carrier frequency. The so-called three-step or
``simple man's" picture describes the underlying physics behind the HHG
phenomenon~\cite{Corkum1993}. In the  first step, occurring about
the maximum of the strong laser electric field, the Coulomb
potential is deformed in such a way that a potential barrier is
formed. Then, the electron is able to tunnel out throughout this
``atomic barrier", and the atom is then ionized. In the second
step, or better to say phase, once the electron is in the
continuum, the electric field of the laser accelerates it.
Naturally, the electron gains energy from the oscillating field,
converting it into a kinetic energy. Consequently, when the
electric field changes its sign, the electron reverses the
direction of its motion and has a certain probability  to
recombine back to the ground state of the remaining ion-core. In
this third step it emits its energy excess as an attosecond burst of coherent
radiation, typically in the XUV or EUV spectral range. In
particular, the maximum emitted harmonic order is about
$n=(I_p+3.17U_p)/\omega_0$, and this result can be explained using
purely classical arguments~\cite{Corkum1993,LewensteinPRA1994}.

One of the main challenges in the production of HHG driven by IR
femtosecond laser fields is the requirement of extra laser
cavities for increasing up the peak power of the oscillator output, given the fact that intensities of the order of 10$^{13}\sim 10^{14}$ W/cm$^2$ are needed for
 the HHG process to happen in atoms. A step forward to mitigate this issue was
 proposed by Kim and co-workers~\cite{Kim2008}.  By focusing a laser pulse of moderate
 intensity, 10$^{11}$~W/cm$^2$, coming directly from a femtosecond oscillator output, onto a
 bow-tie-shaped gold nano-structure array, an enhancement of about $20$~dB of the laser peak intensity on each of the elements was obtained.
 When Argon atomic gas was injected in the vicinity of each nano-structure, high-order harmonic emission
of the fundamental frequency laser-beam was
observed~\cite{Kim2008}.

One should stress, however, that the experimental outcomes of
Kim's experiment are controversial (see
e.g.~\cite{RopersComm,KimReply,RopersNatPhys}); nevertheless, they
have stimulated incessant and promising theoretical activities.
The pioneering theoretical works performed on HHG driven by
plasmonic fields have confirmed two main facts, namely: (i) an
enhancement of the emitted harmonics signal, and (ii) a large
extension in the harmonic cutoff. These two features are mainly
due to the spatial variation at a nanometer scale of the laser
electric field along the laser polarization axis (see
e.g.~\cite{Husakou2011,Ciappina2012,Yavuz2012,CiappiOptExp,JosePRL2013}).

Most of the approaches  to  model HHG, both driven by conventional
and spatial inhomogeneous fields, are based on the hypothesis that a
single active electron (SAE) approximation is good enough to describe
the harmonic emission. Thereby, those pictures neglect
electron-electron interactions in the atomic systems commonly used
to produce high harmonics, such as He, Ar, Xe,
etc.~\cite{Ferray1988}. Nevertheless, studies of HHG considering
two- and multi-electron effects have been performed by several
authors~\cite{ASanpera1,Grobe1993HHG,Koval2007,Bandrauk2005,ShiLin2013,Santra2006,Prager2001}.
From these contributions one could  conclude that depending on the
atomic target properties and the laser frequency-intensity regime,
multi-electron effects could play an important role in  HHG~\cite{Grobe1993HHG,ASanpera1,Koval2007}. 
We should mention,
however, that all these theoretical approaches have been developed
for spatial homogeneous fields and that, to the best of our
knowledge, studies of HHG in two-electron systems driven by
spatially inhomogeneous fields have not been reported yet.


In this Letter we propose plasmonic fields as a tool to study
multi-electron effects in HHG from two-electron systems. We
focus our investigations on the study of the two-electron negative hydrogen ion
(H$^-$) and the helium atom (He). By comparing the numerical
solutions of the reduced 1D$\times$1D-TDSE for both the two-active
electron (TAE), and the SAE models, we
can trace out the analogies and differences in the HHG process
from these two atomic systems, a priori very similar in their
intrinsic structure. The interpretation of our numerical results
renders on a semiclassical approach based on the time-frequency analysis  of
the quantum outcomes, and the classical integration of Newton's
equation of motion.



The 1D models of both H$^-$ and He are described in
Refs.~\cite{Grobe1993HHG} and~\cite{ASanpera1,MLein2000},
respectively, and we shall thus present here only a brief summary.
The HHG spectrum is computed by Fourier transforming the dipole
acceleration $\langle a_d(t) \rangle$; for the TAE model it is
thus mandatory to calculate the two-electron wave function
$\Psi(z_1,z_2,t)$, while for the SAE model the one-electron wave
function $\Psi(z,t)$ is sufficient.  For this aim we numerically
 integrate the reduced 1D-TDSE for both models. In particular, in the case
of the SAE approximation only the outer-electron  is
considered~\cite{ASanpera1,ASanpera2}.

The Hamiltonian $H$ of our two-electron systems can be written in the length gauge as:
\begin{eqnarray}
H=\frac{1}{2}\sum_{j=1}^2 [p_j^2 + V(z_j)] + V_{ee}(z_1,z_2) + V_{\rm int}(t),
\label{Ham2e}
\end{eqnarray}
where $V(z_j)=-\frac{Z}{\sqrt{z_j^2+a}}$ and
$V_{ee}(z_1,z_2)=\frac{1}{\sqrt{(z_1-z_2)^2+b}}$ are the $j$-th
nucleus-electron soft-core Coulomb attractive potential, and the
electron-electron soft-core Coulomb repulsion interaction for our
two-electron system, respectively. Note that $V_{\rm
int}(t)=\sum_{j=1}^2 z_j(1+\frac{\epsilon}{2}z_j)E_{\rm
h}(t)$ defines the coupling of each of the two electrons with the
plasmonic field in the dipole approximation for a linearly
polarized laser field in the $z$-axis. The parameter $\epsilon$
denotes the inhomogeneity strength of the plasmonic field, and has
units of inverse length (for more details see,
e.g.~\cite{Husakou2011,Ciappina2012}). $E_{\rm h}(t)$, is a
spatially homogeneous, or {\it conventional} laser electric field
defined according to $E_{\rm h}(t)=\mathcal{E}_0f(t)\sin(\omega_0
t+\varphi_0)$, where $f(t)$ denotes the pulse envelope and
$\varphi_0$ the carrier envelope phase (CEP).

For completeness we present also the one-electron Hamiltonian:
\begin{eqnarray}
H=\frac{1}{2}p^2 + V(z) + V_{\rm int}(t),
\label{Ham1e}
\end{eqnarray}
where a short-range P\"oschl-Teller potential (PTP)
 $V(z) = -\frac{V_0}{2}\mu^2{\rm sech}^2(\mu z)$~\cite{Poschl33,ChaconCiappinaPeralta} is employed to model the outer-electron of our H$^-$ system and now $V_{\rm int}(t)=z(1+\frac{\epsilon}{2}z)E_{\rm h}(t)$.

In order to compute the two-electron ground state of H$^-$, we set
the soft-core parameters $a=b=1$~a.u., and the nuclear charge
$Z=1$. Then, by imaginary time-propagation with a time step of
$\delta t=-0.03i$~a.u., we integrate the laser-free Schr\"odinger
equation and obtain the ground state wave function. This
calculation yields a binding eigenenergy of ${E}_0=-0.73$~a.u. In
our model for H$^-$, the $I_p$ of the inner-electron is
$I_{p}^{(\rm i)}=0.66$~a.u., thereby the corresponding $I_p$ of
the outer-electron will be about $I_p^{\rm
(o)}\sim0.1$~a.u.~\cite{Grobe1993HHG,RGrobe1993}

For the SAE model, and in order to mimic the $I_p$ of the
outer-electron $I_p^{\rm (o)}$ of the H$^-$, we chose for  the PTP
the screening parameter  $\mu=1$, and $V_0=0.75$~a.u. The
numerical solution of the 1D-TDSE for the TAE and SAE models is
performed using the Split-Spectral Operator method (for more
details see e.g.~\cite{SplitOperator, FFTW,Qfishbowl}).

Hence, to compute the HHG spectra, we first integrate numerically
the 1D-TDSE for both models, and obtain the so-called dipole
acceleration $\langle
a_d(t)\rangle$~\cite{Grobe1993HHG,ASanpera1}, which is the second
order time derivative of electron position in the SAE model, and
the sum of two such terms in the TAE case. The two-electron
position grid space lengths are $L_1=L_2=2500$~a.u., with steps of
$\delta z_1=\delta z_2=0.25$~a.u., respectively. The same grid
parameters are used in case of the SAE approach, but only along a
one-dimensional line. Afterwards, the emitted harmonic yield is
computed by Fourier transforming the $\langle a_d(t)\rangle$,
i.e.~$I_{\rm HHG}=|\mathcal{FT}[\langle a_d(t)\rangle]|^2$. 
The emitted harmonic yields obtained within the TAE model were divided by a factor of 4, in order to take into account the two electrons of the atomic system. In this way a direct comparison with the SAE results can be made. 

In addition to the quantum models, we have numerically solved the Newton's equation $\dot{p}_z(t)=-[1+\epsilon z(t)]E_{\rm h}(t)$,
to compute the classical highest electron energy at recollision $E_{\rm max}$.  We note that the spatial inhomogenous electric field introduces substantial changes in the electron trajectory $z(t)$ (see e.g.~\cite{JosePRL2013,Ciappina2012}). For
conventional fields, i.e.~when $\epsilon=0$, this maximum energy at recollison is given by the usual expression
$E_{\rm max}=3.17 U_p$~\cite{Corkum1993}.
The classical calculation of $E_{\rm max}$ allows us to estimate the maximum harmonic order
 $n_1=({I_p^{(\rm o)}+E_{\rm max}})/{\omega_0}$ of the HHG process driven by
 the inhomogeneous field. To distinguish between the conventional cutoff and the cutoff for the spatial inhomogeneous
 field, we shall denote them as $n_1$ and $n_1'$, respectively.

\begin{figure}[h!]
\includegraphics[width=.238\textwidth]{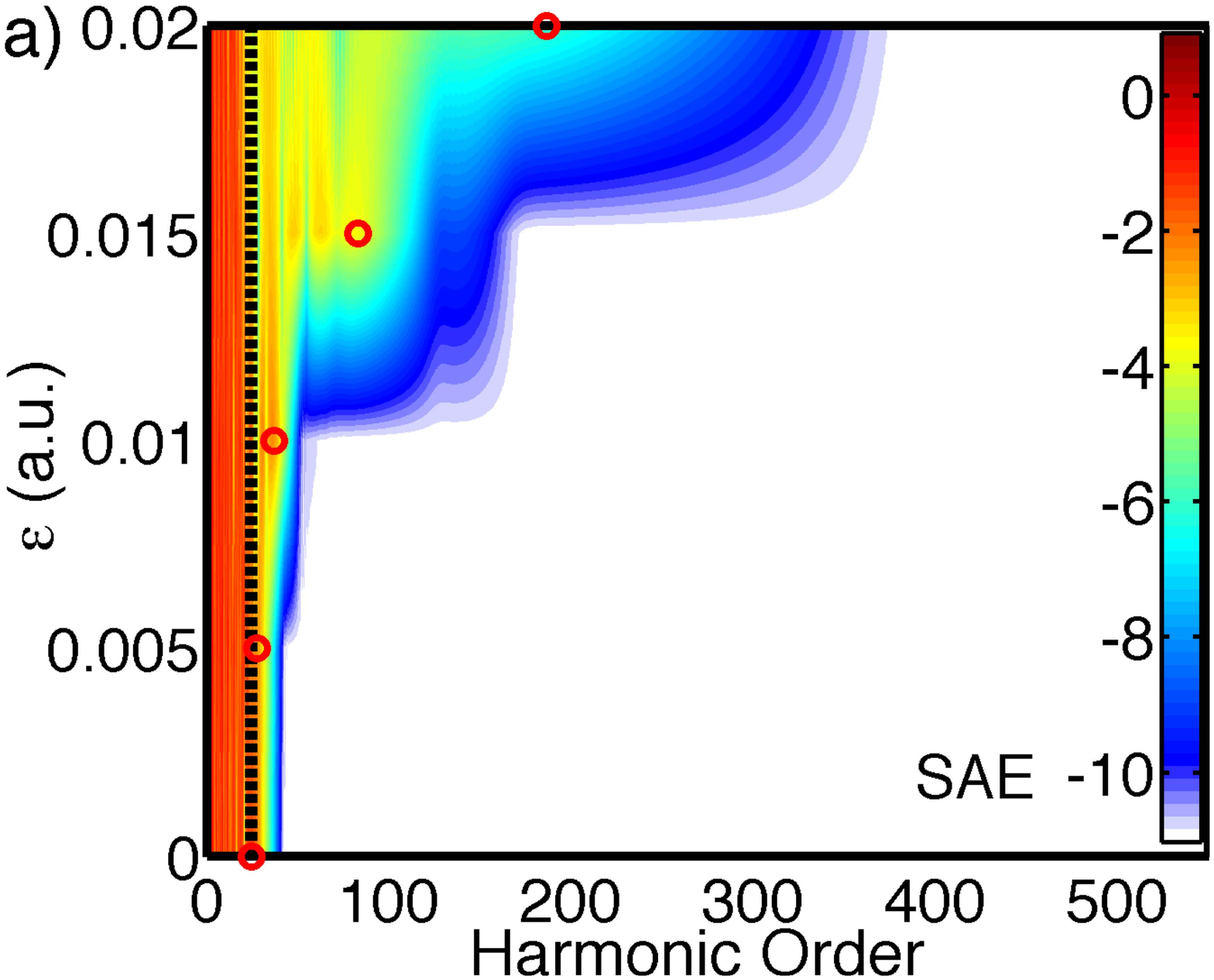}
\includegraphics[width=.238\textwidth]{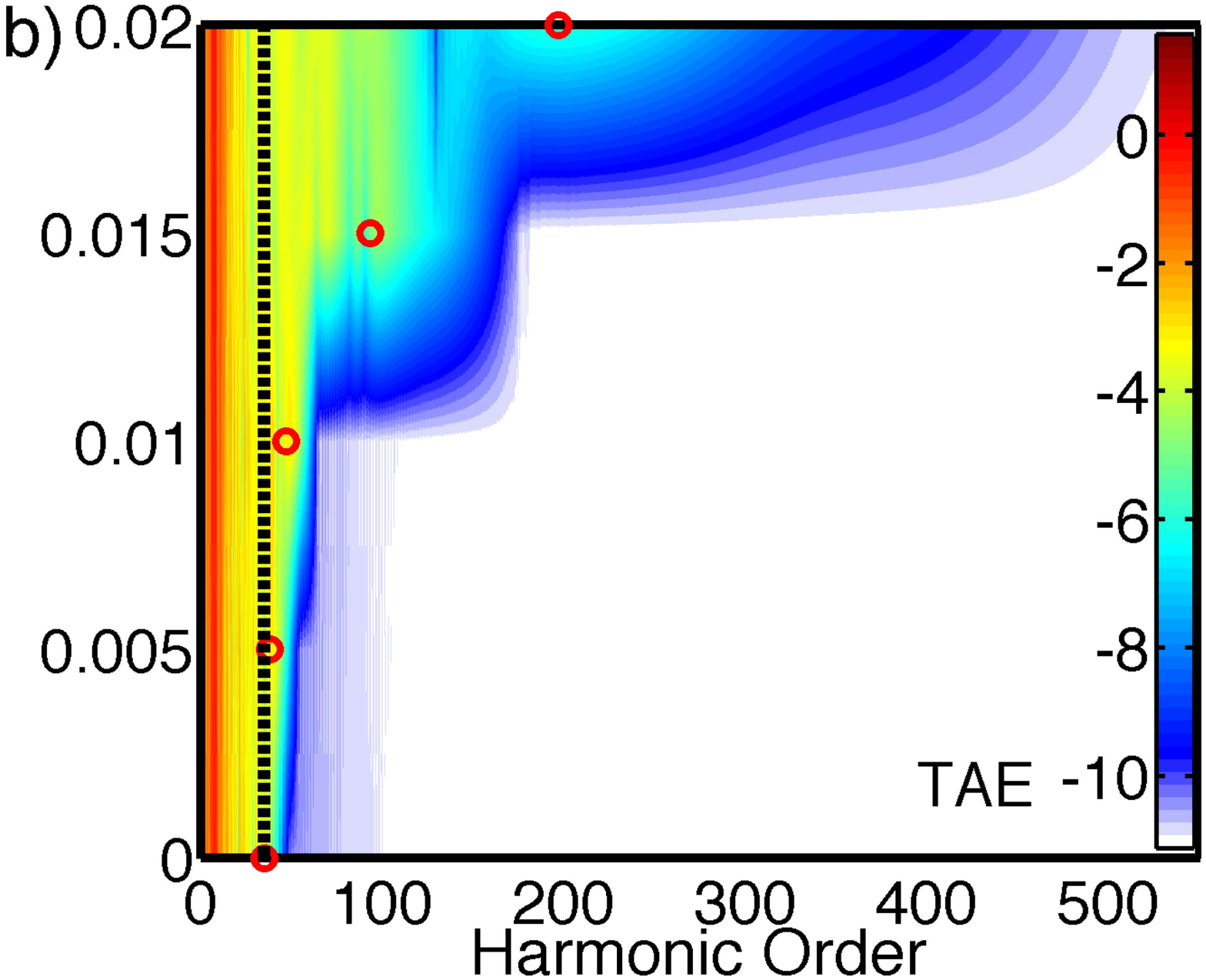}
\caption{\label{Fig1} (color online) HHG yield of H$^{-}$ driven by an inhomogeneous field
in ``logarithm scale" for (a) the SAE and (b) the TAE approximations, respectively.
Black vertical dashed lines denote the cutoff for the SAE conventional field $n_1$ (a) and the
TAE $n_2$ (b). Red circles are the calculated classical cutoffs as a function of $\epsilon$ according to our definition of $n_1'$ and $n_2'$ (see the text for more details). The IR laser pulse parameters are: peak intensity
 $I_0= 2.0\times10^{14}$~W$\cdot$cm$^{-2}$, $\omega_0= 0.057$~a.u. (photon energy 1.55 eV),  $\varphi_{0}= 0$~rad., and $f(t)=\sin^2(\frac{\omega_0t}{2N})$ with 3 total cycles (the corresponding FWHM is 2.6 fs). }
\end{figure}

Fig.~\ref{Fig1} shows the HHG spectra of the H$^-$ as a function
of the inhomogeneity degree, governed by $\epsilon$, for both the
SAE (Fig.~1(a)) and TAE (Fig.~1(b)) models. Clearly, as $\epsilon$
increases, noticeable discrepancies in the harmonic emission
structure and the cutoff are observed between the models. In case
of SAE approach depicted in Fig.~\ref{Fig1}(a), the classical
cutoff, $n_1'$, is in very good agreement with the maximum energy
of the emitted photon. On the contrary, for the TAE model, the
classical predicted maximum harmonic order, $n_1'$, is unable to
match the cutoff obtained quantum mechanically, even for the case
of $\epsilon=0$. Hence, we denote this ``new" cutoff for the H$^-$
TAE model by $n_2$. Logically, a natural question arises: where
this clear disagreement between the SAE and TAE model comes from
(note that the disparity is clearly visible for both
conventional and spatial inhomogeneous electric fields cases).

\begin{figure}[h!]
\includegraphics[width=.237\textwidth]{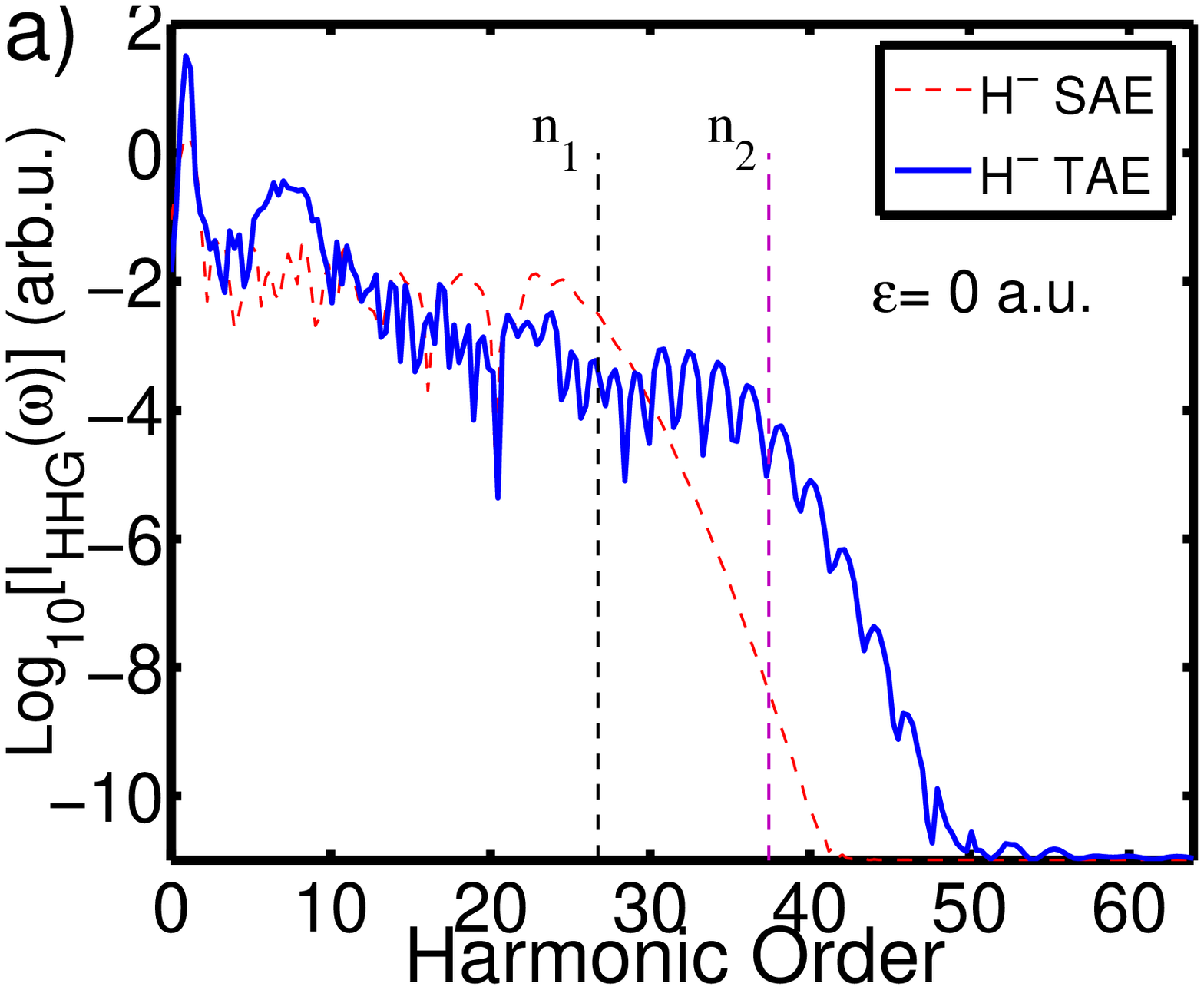}
\includegraphics[width=.237\textwidth]{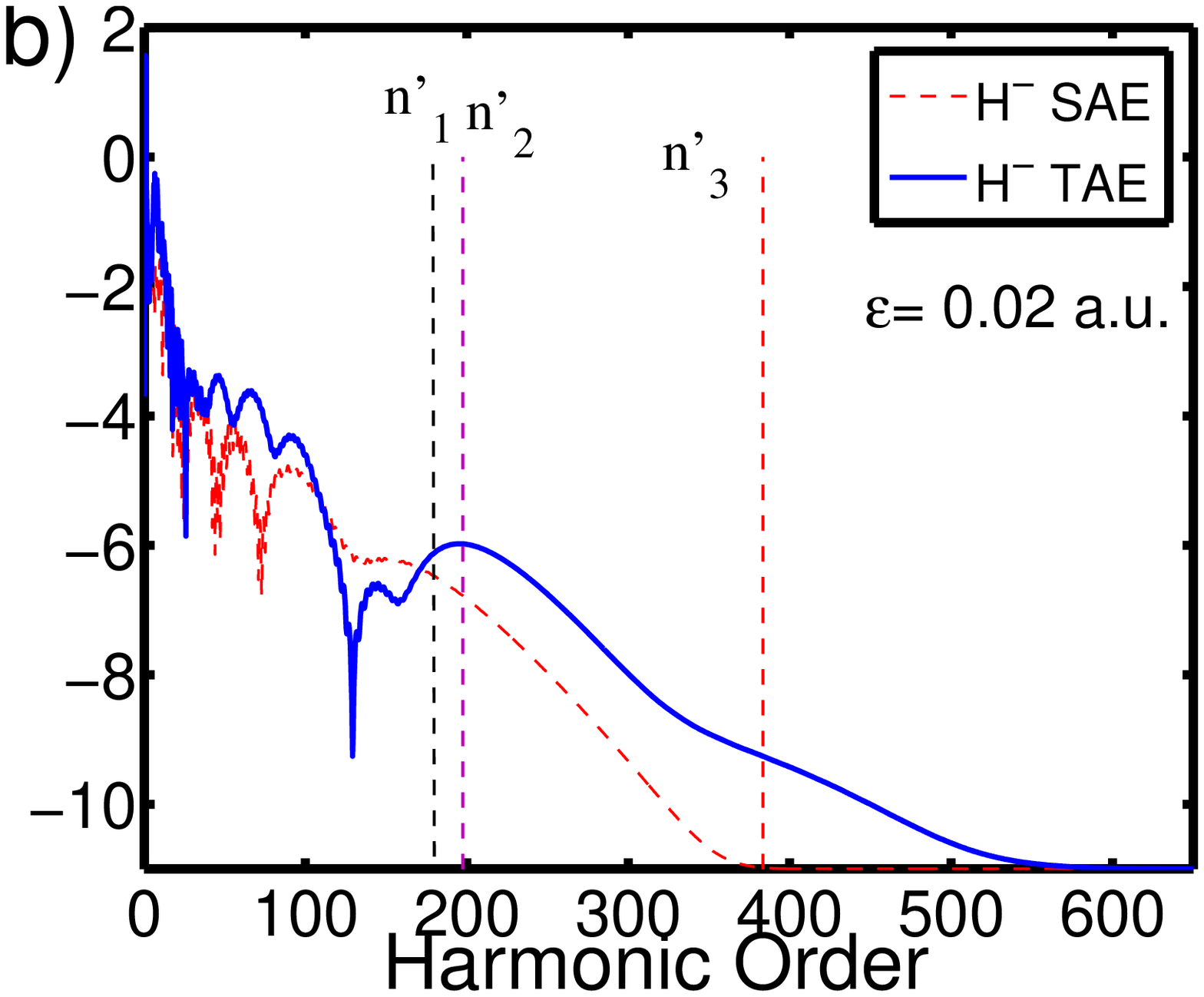}
\caption{\label{Fig1A} (color online) (a) conventional and (b) inhomogeneous emitted HHG
spectra of H$^{-}$ for our SAE and TAE models. In the HHG driven by a conventional field,
the classical cutoff energy for the SAE and TAE calculations are denoted by
$n_1$ (black vertical dashed line) and $n_2$ (violet vertical dashed line), respectively.
The $n_1'$, $n_2'$ and $n_3'$ (red vertical dashed line) denote the
cutoff for the SAE, TAE and the double non-sequential two-electron re-combination mechanism
(see the text), respectively.}
\end{figure}

In order to address the above question, in Figs.~\ref{Fig1A}(a)
and ~\ref{Fig1A}(b) we compare the emitted HHG spectra for two
specific cases: the conventional case $\epsilon=0$ and an
inhomogeneous case with $\epsilon=0.02$~a.u., respectively.

On the one hand, for conventional fields and in the case of
two-electron systems, where double-electron ionization
contributions are not relevant, one can expect that the cutoffs
$n_1$ and $n_2$ coincide~\cite{ASanpera1}. The results depicted in
Fig.~\ref{Fig1A}(a) clearly shows this is not the case: an extra
extension in the $I_{\rm HHG}^{\rm(TAE)}(\omega)$ cutoff with
respect to the SAE model, is found. According to Lappas {\it et
al}.~\cite{ASanpera1} possible inner-electron contributions to the
harmonic spectrum should extend the cutoff by an extra amount of
$n_{\rm{shift}}=(I_p^{\rm(i)}-I_p^{\rm(
o)})/\omega_0$~\cite{ASanpera1,Grobe1993HHG,Koval2007}. Hence, we
argue that the main mechanism behind this HHG extension is the
sequential double-electron ionization of the outer- and
inner-electrons, and re-collision of both of
them~\cite{Koval2007}. This leads to a cutoff given by
$n_2=n_1+n_{\rm shift}$ which is in reasonable agreement with the
HHG spectrum computed by our TAE model.

On the other hand, Fig.~\ref{Fig1A}(b) shows the emitted HHG
spectra driven by a spatial inhomogeneous field for both the SAE
and TAE models. Firstly, and similarly to the conventional field
case, a structural difference between the models is observed in
the HHG spectra, which comes from the different events
of the electron re-combinations. Furthermore, the harmonic cutoff
predicted by the SAE model is well reproduced by $n_1'$. Secondly,
a large high harmonic energy ``cutoff" appears in the case of the
computed HHG by means of the TAE model. That ``cutoff" cannot be explained by
$n_2'$, even when the shift $n_{\rm shift}$ is included. This
means that another {\em mechanism} is responsible for this
extension. One candidate to explain this extension could be  a
double non-sequential two-electron re-combination
event~\cite{Koval2007}. In order to probe this hypothesis, from
classical calculations we can infer that the two electrons will
re-combine to the ground state with maximum energies $E_{\rm
max}^{\rm (1st)}$ and $E_{\rm max}^{\rm (2nd)}$, respectively, and
then the emitted harmonic order cutoff could be obtained
from~\cite{Koval2007}:
\begin{eqnarray}
n_3'= (E_{\rm max}^{\rm (1st)}+E_{\rm max}^{\rm (2nd)}+I_p^{\rm(i)}+I_p^{\rm(o)})/\omega_0, \label{eqn:EqN3}
\end{eqnarray}
where, $E_{\rm max}^{\rm (1st)}$ and $E_{\rm max}^{\rm (2nd)}$
denote the maxima first (inner-electron) and second (outer-electron)
re-collision energies, respectively. From Fig.~\ref{Fig1A}(b) we
observe that the predictions of Eq.~(\ref{eqn:EqN3}) are in
excellent agreement with our TAE model calculations (see the
vertical dashed red line). Next, we shall explain in detail the
underlying  physics of Eq.~(\ref{eqn:EqN3}).

\begin{figure}[h!]
\includegraphics[width=.2354\textwidth]{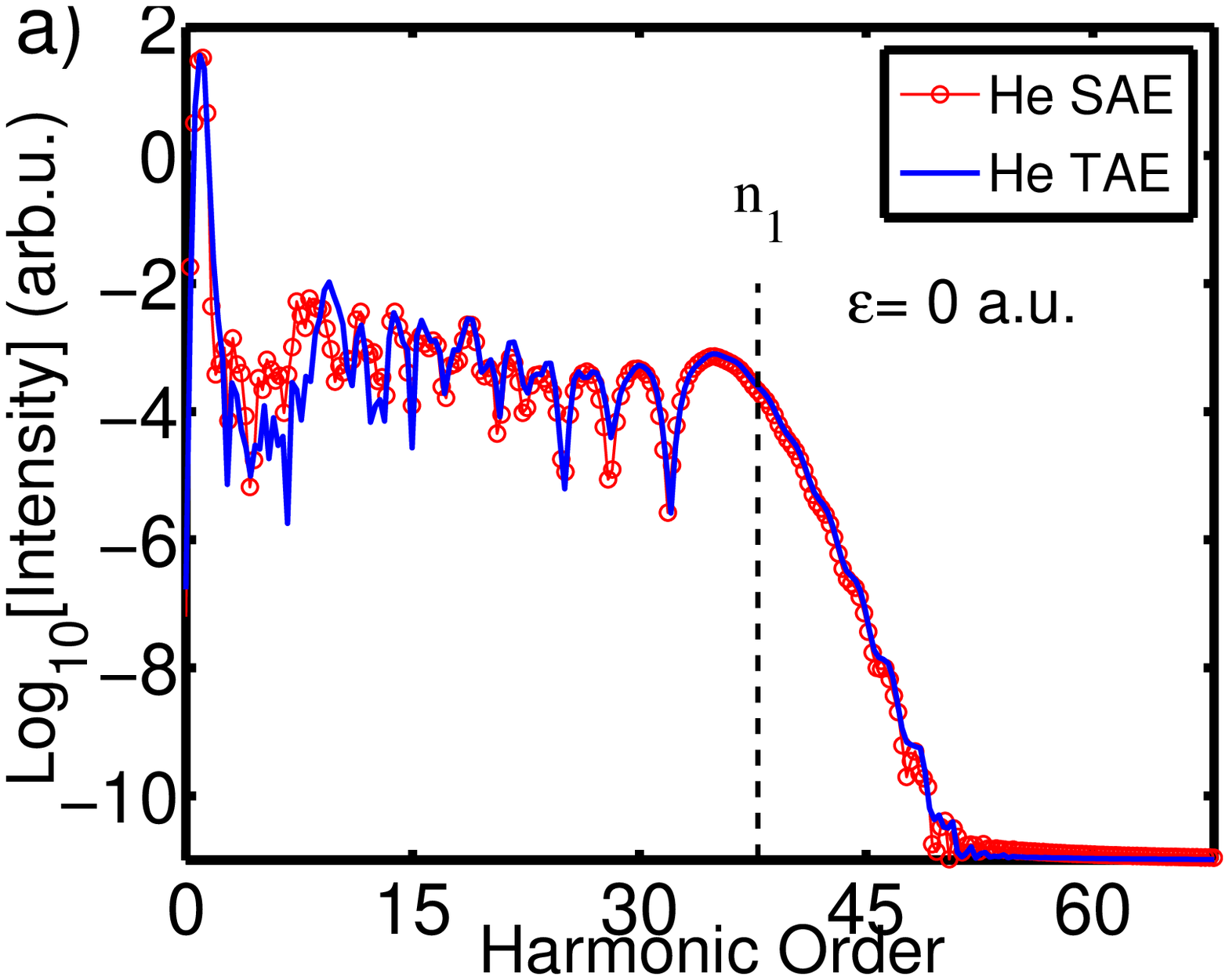}
\includegraphics[width=.2354\textwidth]{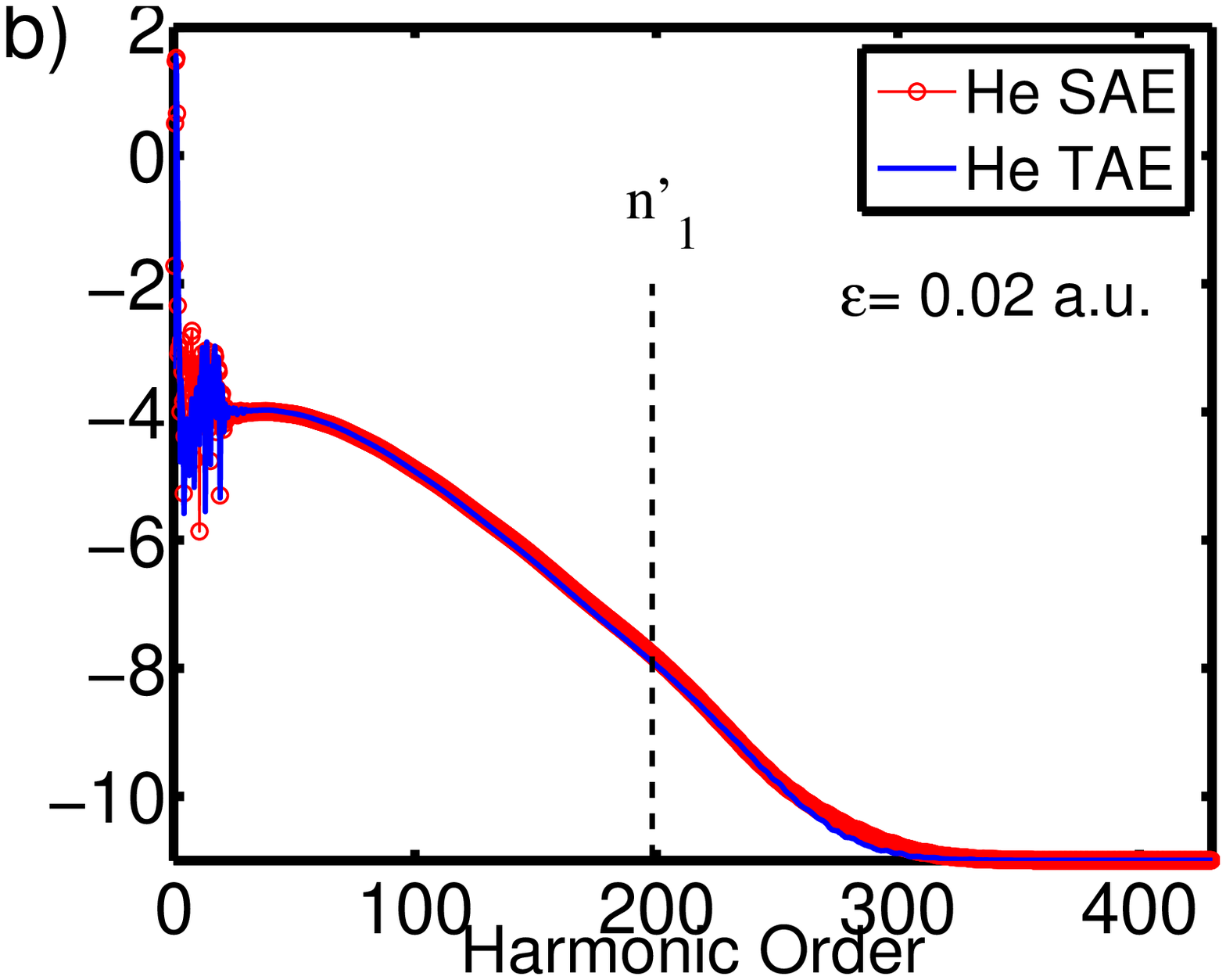}
\caption{\label{Fig3} (color online) HHG for the He atom. The HHG spectra of (a) and (b) are computed with the same parameters as in Fig.~\ref{Fig1A}. Black dashed vertical lines
denote the $n_1$ conventional field cutoff for the outer-electron (a),
and $n_1'$ the inhomogeneous field cutoff (b). }
\end{figure}

Pursuing to find a broader panorama about the behavior of
two-electron systems we have computed the HHG spectra of the He
atom for both conventional and spatially inhomogeneous fields. The
same SAE and TAE models of He described
in~\cite{ASanpera1,MLein2000} have been implemented.  Similarly to
the H$^-$, for the SAE approach we have integrated in a 1D-TDSE
the outer-electron of He. In such a case the electron is simulated
by the long-range soft-core Coulomb potential described
in~\cite{ASanpera1}. The outer-electron $I_p$ of our He model is
about $I_p^{\rm(o)}=0.73$~a.u. The results of the HHG spectra for
both the conventional and spatial inhomogeneous fields are
depicted in Figs.~\ref{Fig3}(a) and \ref{Fig3}(b), respectively.
Clearly, we can observe that both approaches are in perfect
agreement for both cases. Slightly reasonable
discrepancies are found for low-order harmonics, a region where
the details of the atomic potential are relevant. In addition,
both the SAE and TAE models' cutoffs are identical, and as a
consequence we  argue that the outer-electron is the main
responsible of HHG emission process in case of He. Note that an
intensity scan of
1$\times$10$^{14}$~--~9$\times$10$^{14}$~W/cm$^2$ in the
calculation of the HHG spectra have been performed for different
values of $\epsilon$. Same degree of agreement between the SAE and
TAE pictures was found. It is remarkable that neither an extra
extension, $n_{\rm shift}$, nor an underestimation on the $n_3'$
cutoff is observed for spatial inhomogeneous fields in the TAE
model of He.

From this last asseveration, we can conclude that single-
and
 double-electron ionization effects play an important role in
the description of the emitted HHG from the H$^-$ driven by both
conventional and inhomogeneous fields. The latter provide a novel
instrumental tool in order to enhance two-electron effects.


\begin{figure}[h!]
\includegraphics[width=.239\textwidth]{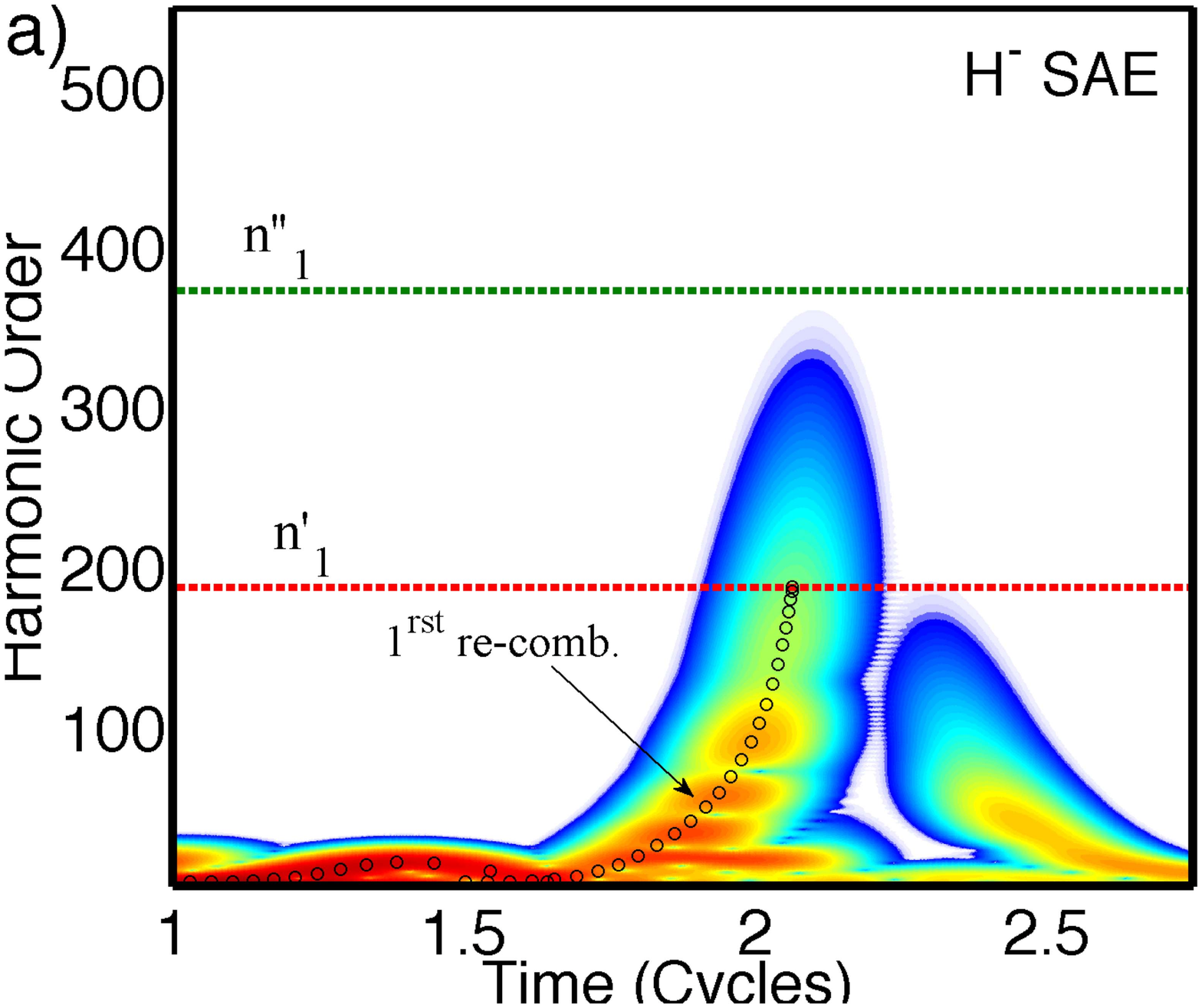}
\includegraphics[width=.238\textwidth]{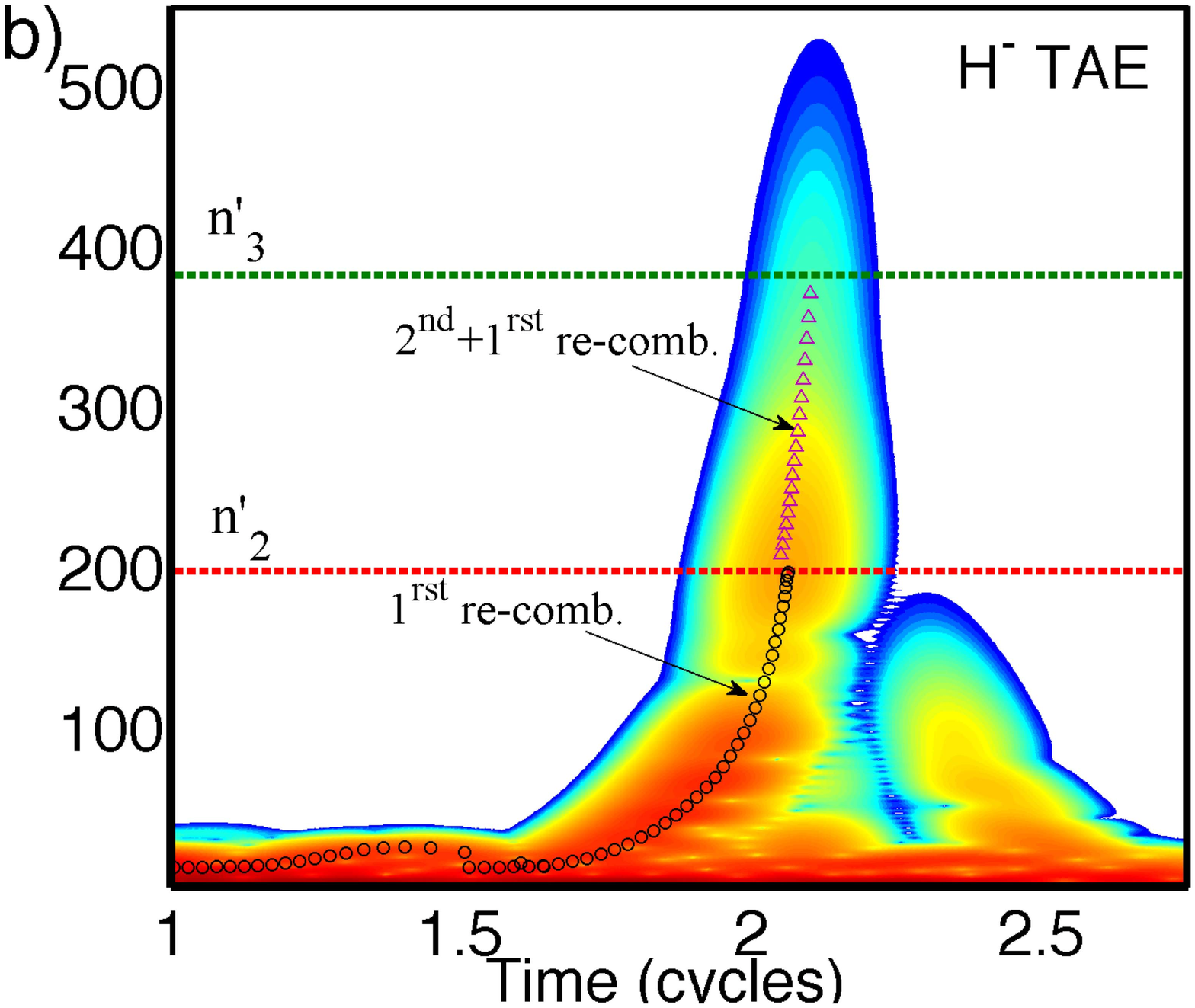}
\caption{\label{Fig4} (color online) Gabor distributions and semiclassical analysis of the HHG
driven by inhomogeneous fields for the SAE (a) and TAE (b) models of H$^-$.
Violet circles depict the ``classical" harmonic emission times considering the first electron re-combination events
 driven by the inhomogeneous field.  In (b) the white triangles are the emitted harmonics
by classically considering the first and the second double non-sequential two-electron re-combination events.
 In (a) the red and green horizontal dashed lines denote the cutoff $n_1'$
 and a modification of $n_1'$ which takes into a count the first and second
 re-combination ($n_1''$) for the SAE. In case of the TAE model (b), the same horizontal
 lines depict the cutoffs denoted by $n_2'$ and $n_3'$ formulae (see the text).}
\end{figure}

In order to further clarify the reasons of the extended HHG
spectra for the H$^-$ we perform a time-analysis of the quantum
mechanical results in terms of the Gabor distribution (for details
see e.g.~\cite{Gabor,ManfredGabor}). In short, information about
the classical electron trajectories can be extracted from the HHG
spectra computed via the TDSE, and compared with pure classical
calculations.

Figure \ref{Fig4} depicts the Gabor distribution obtained from the
quantum mechanically computed HHG spectra of the H$^-$ system
driven by an inhomogeneous field for both the SAE
(Fig.~\ref{Fig4}(a)) and TAE~(\ref{Fig4}(b)) models,
respectively. The emitted harmonics calculated by the first and
second classical
 re-collision electron-trajectories are also depicted.
Excellent agreement between the Gabor synthesis and the classical
calculations is found in case of the SAE picture. Note that only
the first re-collisions are considered in the classical
calculations shown in Fig.~\ref{Fig4}(a). However, when the TAE
approach is employed, the first electron re-collision events are
not sufficient to reproduce the whole range of emitted harmonics
described by the Gabor time-frequency decomposition. Nevertheless,
if one considers a second re-collision event in the classical
approach, the maximum harmonic emission of Gabor's distribution is
then perfectly reproduced.

 These observations suggest that the main mechanism behind the HHG spectra emitted from our two-electron H$^-$ model
 driven by a spatially inhomogeneous field can be summarized as follows: (i) the outer-electron is ionized via tunneling
 about one of the
maxima of the IR laser pulse. Then, after a while, around the
second consecutive maxima, the inner-electron is liberated; (ii)
as the single-electron and double-electron ionization
probabilities are large, the outer-electron has a high chance to
make a second re-collision event together with the first
re-collision of
 inner-electron and {\em both} at the {\em same} re-combination time.  This analysis indicates the
 emission of a ``maximum" high-harmonic photon of order $n_3$ (see Eq.~(\ref{eqn:EqN3})). Then,
 the emitted photon at this particular time will be the sum of the first and second re-collision
 energies of the two electrons with their respective ground states. This picture is in perfect agreement with the Gabor
 distribution results, extracted from the quantum mechanical models. Furthermore, it is the spatially inhomogeneous character
 of the laser electric field which is the main responsible of the increase of the probability of this peculiar mechanism.

Additionally, the Gabor distribution of Fig.~\ref{Fig4}(a) helps us to disentangle
if the inner- and outer-electron are
re-combining with the remaining ion-core at the same re-collision
time,  or if it is only a first and second re-combination of a
single-electron process. As it
is clearly shown, there is not emitted harmonic signal at
$n_1''=(E_{\rm max}^{\rm(1st)}+E_{\rm max}^{\rm(2nd)}+I_p^{(\rm
o)})/\omega_0$. Then, it is not possible that the cutoff extension
in the HHG spectra obtained from the TAE model comes from a
single-electron ionization event followed by a first and second
re-combinations event of this unique electron.

Hence, we believe we have collected convincing arguments, based
both on quantum mechanical and classical analysis, that a new
mechanism, a double non-sequential  two-electron re-combination is
the main responsible of the extension of the HHG spectra of the
H$^-$ when a spatial inhomogeneous plasmonic field is used to
drive the process.

Summarizing, we performed two-electron calculations of HHG driven
by spatial homogeneous and inhomogeneous fields. We used as test
systems H$^{-}$ and He. By the numerical solution of 1D-TDSE
models in the two- and single-active-electron approximations,
supported by  a classical analysis of the electron trajectories,
we demonstrated that an extra extension in the harmonic cutoff was
found in case of the TAE model for H$^{-}$, which cannot be
explained within the SAE framework. After a comprehensive analysis using
complementary tools, we concluded that a new mechanism is the main
responsible of this extension. One of the main advantages to use
plasmonic fields as ``probes" is the low incoming intensity needed
in order to observe this effect, considering that the plasmonic
nano-structures act as light amplifiers. In addition, as we have
shown, the results strongly depend on the atomic system employed.
In particular, the H$^-$ was chosen for simplicity, but we
consider similar effects and results could be found for highly
correlated negative ion systems, such as the Alkali negative
ions~\cite{McLean}.

\begin{acknowledgments}
 A. C. and M. L. acknowledge MINECO grant FOQUS(FIS2013-46768-P), AGAUR  grant 2014 SGR
874,  Fundacio Cellex Barcelona, and ERC AdG OSYRIS.
 We also acknowledge the support from the EC's Seventh Framework Programme LASERLAB-EUROPE III
 (grant agreement 284464) and the Ministerio de Econom\'ia y Competitividad of 
 Spain (FURIAM project FIS2013-47741-R).
\end{acknowledgments}



\end{document}